\documentclass[superscriptaddress,showpacs,aps,twocolumn,prb,floatfix]{revtex4}

\usepackage{bm,amsmath,amssymb}
\usepackage{graphicx}
\usepackage{color}
\begin{document}

\title{Spin and orbital ordering in bilayer Sr$_3$Cr$_2$O$_7$}

\author{Armando A. Aligia}
\affiliation{Centro At\'omico Bariloche and Instituto Balseiro, CNEA, 8400 S. 
C. de Bariloche, Argentina}
\affiliation{Consejo Nacional de Investigaciones Cient\'{\i}ficas 
y T\'ecnicas (CONICET), Buenos Aires, Argentina}

\author{Christian Helman}
\affiliation{Centro At\'omico Bariloche and Instituto Balseiro, CNEA, 8400 S. 
C. de Bariloche, Argentina}

\date{\today }

\begin{abstract}
Using maximally localized Wannier functions obtained from DFT calculations, we 
derive an effective Hubbard Hamiltonian for a bilayer of Sr$_3$Cr$_2$O$_7$, the 
$n=2$ member of the Ruddlesden-Popper  Sr$_{n+1}$Cr$_n$O$_{3n+1}$ system. 
The model consists of effective $t_{2g}$ orbitals of Cr in two square lattices, 
one above the other. The model is further reduced at low energies and two 
electrons per site, to an effective Kugel-Khomskii Hamiltonian that describes 
interacting spins 1 and pseudospins 1/2 at each site describing spin and 
orbitals degrees of freedom respectively. 
We solve this Hamiltonian at zero temperature using pseudospin bond operators 
and spin waves. 
Our results confirm a previous experimental and theoretical study that proposes 
spin ordering antiferromagnetic in the planes and ferromagnetic between planes, 
while pseudospins form vertical singlets, although the interplane separation is 
larger than the nearest-neighbor distance in the plane. 
We explain the physics behind this rather unexpected behavior. 
\end{abstract}

\maketitle

\section{Introduction}
\label{intro}

Some decades ago, Kugel and Khomskii studied theoretically the interplay 
between orbital and spin degrees of freedom in compounds like KCuF$_3$ and 
K$_2$CuF$_4$.\cite{kugel} 
They showed that in these compounds the $e_g$ orbital degrees of freedom 
(leaving the hole in the 3d$^9$ configuration of Cu in the orbital with symmetry 
either $x^2-y^2$ or $3z^2-r^2$) can be described by a pseudospin, and these 
pseudospins interact between them and with the spins of the Cu ions, in such a 
way that the preferred ordering is antiferromagnetic for the pseudospins and the 
spin ordering is ferromagnetic in the $ab$ plane and antiferromagnetic in the 
$c$ direction.
The staggering ordering of the orbitals leads to a staggering of quadrupolar 
distorted CuF$_4$ units in the $ab$ planes, as expected from any electron-phonon 
interaction.\cite{liech} 

The interest in systems with spin and orbital degrees of freedom was rising 
over the years. See for example Refs. 
\onlinecite{liech,mizo,Feiner,ling,Ulrich,Horsch,Fang,ruo,Sugai,Pavarini,Manaka,bruce,van,Stingl,
Oles,kov,fepc,fepc2,giova,srcro}.
A rich physics has been observed in the $n=2$ members of the Ruddlesden-Popper 
series of the form $A_{n+1}B_nC_{3n+1}$, where $B$ denote transition-metal atoms 
that form two square lattices in the $xy$ plane, one displaced with respect to 
the other in the $z$ direction.\cite{ling,Manaka,Stingl,srcro} 
In the layered colossal magnetoresistance manganite 
La$_{2-2x}$Sr$_{1+2x}$Mn$_2$O$_7$, different spin and orbital orderings are 
observed as $x$ is varied indicating that $e_g$ orbital polarization is the 
driving force behind spin ordering.\cite{ling}
Particularly interesting is K$_3$Cr$_2$O$_7$, where distortions reveal 
antiferromagnetic orbital ordering, while the system presents a spin gap due to 
spin dimers in the $z$ direction.\cite{Manaka} 
This is an exotic ordering involving spins and pseudospins both of magnitude 
1/2. 
In Sr$_3$Ru$_2$O$_7$, an applied magnetic field induces domains with distorted 
lattice parameters likely related to orbital ordering.\cite{Stingl}

Recently the compound Sr$_3$Cr$_2$O$_7$ has been studied by several experimental 
and theoretical techniques.\cite{srcro} 
The resistivity and specific heat measurements indicate that the system is 
insulating. 
The calculations using density functional theory (DFT) with LDA+U 
approximation indicate that occupancy of Cr is consistent with an oxidation 
state Cr$^{+4}$ (two electrons in the 3d shell) 
and there is an orbital degeneracy between $d^1_{xy}d^1_{xz}$ and 
$d^1_{xy}d^1_{yz}$. 
Hund rules favor a total spin $S=1$. 
There is a magnetic transition at 210 K with a huge total entropy change near 
$R\ln(6)$ indicating a simultaneous spin and orbital ordering. 
The magnetic structure observed by neutron diffraction is consistent with 
the DFT results and correspond to antiferromagnetic alignment between 
nearest-neighbors in the plane and ferromagnetic between planes. 
Orbital ordering is usually not detected in DFT with LDA and derived potentials 
due to the difficulties of these techniques to obtain orbital 
polarization.\cite{liech,note} LDA+U is able to capture orbital ordering,\cite{liech} 
but not singlet ordering because it is based on a single Slater determinant.
The LDA+U calculations of Ref. \onlinecite{srcro} do not detect any orbital ordering.
Nevertheless, the absence of observable distortions that accompany the orbital 
ordering (as in KCuF$_3$ and K$_2$CuF$_4$ mentioned above) is consistent with 
the formation of vertical singlet orbital dimers, so that quantum fluctuations 
destroy long-range pseudospin ordering. 
This spin/pseudospin state has some analogy to the case of K$_3$Cu$_2$O$_7$ 
(with both spin and pseudospin $S=1/2$) mentioned in the previous paragraph 
with spin and pseudospin (orbital) degrees of freedom interchanged.

The authors have also derived a Kugel-Khomskii Hamiltonian from a multiband 
Hubbard model containing the relevant orbitals of Cr and O.\cite{srcro}. 
Solving this Hamiltonian by Lanczos in a cluster containing eight sites, 
they found that effectively the ground state corresponds to the observed 
magnetic ordering and vertical pseudospin singlets if $t^{\prime}_z/t^{\prime}_p 
> 0.85$, where $t^{\prime}_z$ ($t^{\prime}_p$) is the hopping between 
nearest-neighbor Cr and O orbitals perpendicular to (in) the CrO$_2$ planes.
Since $t^{\prime}_z$ is expected to be smaller than $t^{\prime}_p$ because of 
the larger distances in the plane, and the size of the cluster is very small, 
further theoretical work is necessary to confirm that the proposed exotic spin 
and orbital ordering is in fact the ground state. Note that for a spin 1/2 
Heisenberg model in a bilayer system (our bilayer system for pseudospins 
only, i.e. $S=0$) with interplane interaction $J$ and vertical one  $J^\prime$, the quantum 
phase transition from a Neel ordered phase to the quantum disordered 
dimer-singlet phase takes place for $J^\prime/J \approx 2.522$,\cite{sand,wang}
and for the three-dimensional extension with two cubic sublattices the 
transition moves to $J^\prime/J \approx 4.84$.\cite{qin}
Therefore naively one would expect a phase with long range pseudospin ordering 
in Sr$_3$Cr$_2$O$_7$ for $t^{\prime}_z \sim t^{\prime}_p$.   

In this work we first construct a tight-binding model for {\em effective} 
$t_{2g}$ orbitals at the Cr sites, using maximally localized Wannier functions 
(MLWFs) and add to it the on-site interactions.\cite{ruo,spli} 
This leads to a three-band Hubbard model. This starting approach is similar to 
that followed by Ogura {\it et al.} to study the possible occurrence of 
superconductivity in hole doped Sr$_3$Cr$_2$O$_7$ and 
Sr$_3$Mo$_2$O$_7$.\cite{superc} 
These effective $t_{2g}$ orbitals are not pure Cr 3d orbitals of $xy$, $xz$, and 
$yz$ symmetry but contain an important admixture with O  orbitals. 
Next we derive a Kugel-Khomskii Hamiltonian for the low-energy subspace of two 
electrons per site, by degenerate perturbation theory in the hopping terms. We 
explain the meaning of the different terms and the expected physics. 
Finally this Hamiltonian is solved for the infinite system, using a combination 
of bond operators and spin waves. 
For the resulting parameters we obtain that the state of singlet dimers and the 
spin ordering proposed in Ref. \onlinecite{srcro} is in fact the ground state.
To destabilize it, one would need to reduce the hopping $t^{\prime}_z$ 
mentioned above by a factor near 1/2.

The paper is organized as follows. In Section \ref{sys} we describe the 
atomic and electronic structures. Section \ref{hopar} describes the 
\textit{ab initio} method used to obtain the effective hoppings 
and on-site energies used in Section \ref{h6b} to derive an effective
multiband model for the system. In Section \ref{hkk} we use this model to derive
an effective Kugel-Khomskii Hamiltonian to describe the spin and orbital
degrees of freedom of the model. This model is solved in Section \ref{spso} 
using a generalized spin-wave theory. Section \ref{sum} contains a summary 
and discussion.

\begin{figure}[t]
  \includegraphics[width=7cm]{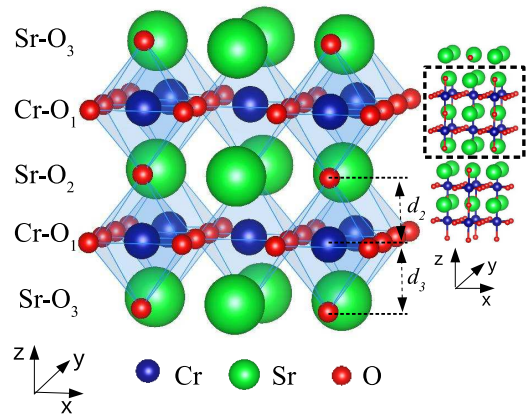}
  \caption{\label{estructura} (Color online) Tetragonal structure of Sr$_3$Cr$_2$O$_7$. The 
stacking is made of three types of layers. The unit cell (shown at the right) 
contains two blocks of five layers shown at the left, the second displaced in 
the $x,y$ direction by $(a/2,a/2)$ with respect to the first one. 
$d_2$ and $d_3$ denote the distances 
Cr-O$_2$ and Cr-O$_3$, respectively. The ratio $d_3/d_2=1.016$.\cite{srcro}   }
\end{figure}

\begin{figure}
 \includegraphics[trim=1cm 1cm 1cm 1cm, clip, scale=0.5, 
keepaspectratio=true]{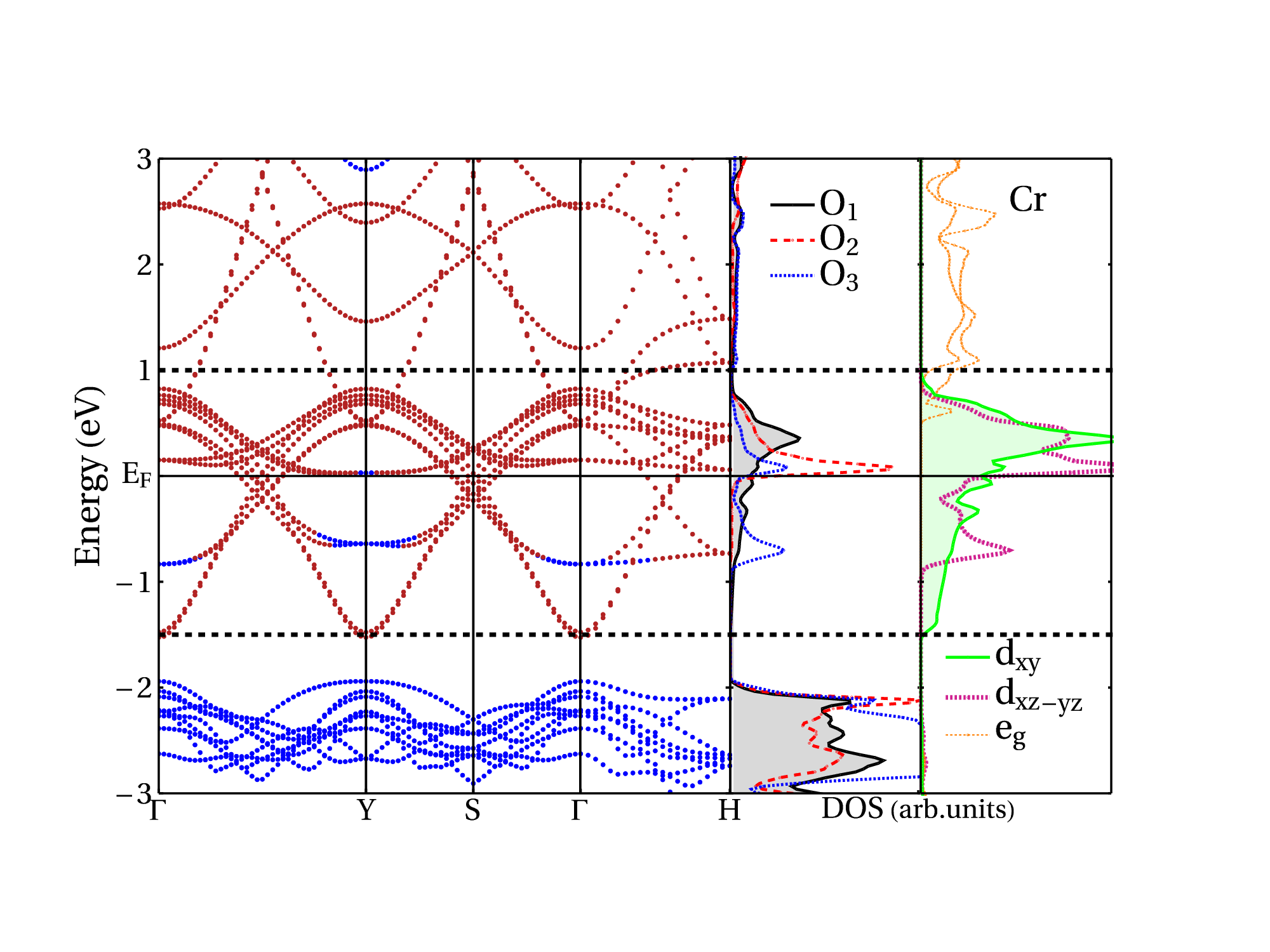}
\caption{\label{bandas} (Color online)
Spin unpolarized band structure of Sr$_3$Cr$_2$O$_7$ 
(left panel)  along 
with atom-projected density of states (right panel). The bands are plotted with 
character in order to show the strong hybridization between Cr and O close to the
Fermi energy. Red means mainly $d$-Cr character and blue  $p$-O character. 
Also shown are the DOS projected on each non-equivalent O atom and each Cr orbital.
The two  horizontal dashed lines delimit the energy window where the 
\textit{wannierization} process takes place.}
\end{figure}

\section{The System}
\label{sys}

As a starting point we calculate the electronic structure of the 
Sr$_3$Cr$_2$O$_7$ system within the framework of DFT. 
The structure has I4/mmm symmetry, which is  tetragonal  with lattice 
parameters $a=3.796$\AA\  and $c=19.846$\AA\ \cite{srcro} and we include the 
asymmetric deformation of the CrO$_6$ octahedra as reported in Ref. 
\onlinecite{srcro}.
We distinguish three types of O atoms according to its bonding role.
As shown in Fig. \ref{estructura}, the structure is a stacking of CrO$_2$ layers 
and SrO layers. We label as O$_1$ the O atoms inside the CrO$_2$ layers,
O$_2$ denote the O atoms in the SrO layers between two CrO$_2$ layers,
and O$_3$ refer to the O atoms of the SrO layers that lie between
another Sr-O$_3$ layer and a CrO$_2$ one.
Two consecutive Sr-O$_3$ layers are displaced by a vector $(a/2,a/2,d)$, where $d$ is
the interlayer Sr-O$_3$ distance,  as can be 
seen in the small scheme at the right of Fig. \ref{estructura}. 
In the CrO$_2$ layers, both, O$_1$ and Cr atoms form a square lattice. 
Each Cr atom has four nearest neighbors that are O$_1$ atoms in the $x,y$ plane and 
two more O neighbors in the $z$ direction,  the one between CrO$_2$ 
layers is type O$_2$ and the other is type O$_3$. Note that the distances between Cr and O$_2$ and 
between Cr and O$_3$ are different, as shown in Fig. \ref{estructura}.

Since we are looking for the hopping parameters, the DFT calculations 
are performed as spin unpolarized, but using both, cell parameters and atomic 
coordinates obtained from the spin-polarized case.
The band structure calculations are done using Wien2k code\cite{wien2k}, with 
precision parameters $R_{MT}.K_{max}=7$, which reads as the product between 
smallest muffin tin sphere radius with the plane-wave cutoff ($K_{max}$). 
The Brillouin zone was sampled with a regular mesh containing 800 irreducible 
points and we use the GGA approximation for the potential of
exchange-correlation. 

The  band structure along with atom-projected density of states are shown
in Fig. \ref{bandas}, where band character is emphasized by color, red means 
strong Cr component and blue stands for O component. 
The electronic structure close to the Fermi energy is dominated by Cr-d 
states, which are split by the tetragonal component of the crystal field 
into $t_{2g}$ and $e_{g}$ states. The Cr-$t_{2g}$ states are the partially 
filled states and share a peak with O$_3$ at -0.7 eV, also they hybridize with 
in plane O$_1$ atoms. Above the Fermi level, the states of Cr share a peak with 
O$_2$ at 0.09 eV.
These characteristics in the DOS expose the bridging character of the O$_2$ 
between next CrO$_2$ layers.
The band structure presents a region around the Fermi level, where the 
$t_{2g}$ states of the Cr prevail.
The $e_g$-type orbitals are located at 0.9 eV above the Fermi level. 

Note that the band structure predicts a metallic state while actually the compound
is an insulator.\cite{srcro} This is corrected when the interactions are included,
as we do in Section \ref{h6b}.

\section{Hopping parameters from MLWFs}
\label{hopar}

The MLWFs approach provides a physically intuitive and also rigorous 
representation of the electronic band structure of a system in an energy region 
of interest, which defines a Hilbert subspace.
Then, the Hamiltonian expressed in the base of MLWFs can be mapped to a tight  
binding-based model which describes the system in the target Hilbert subspace.
These MLWFs can be derived from Bloch states of a DFT calculation by the so  
called \textit{wannierization} process as implemented in the Wannier90 
code\cite{wannier90}.
However, the input of Wannier90  requires the overlap matrices and projections to the 
Hilbert subspace, so we use the Wien2Wannier\cite{wien2wannier} routine as 
interface among them.

The target Hilbert subspace is chosen in order to describe the Cr  
$d$-orbitals near the Fermi energy, which are responsible of the magnetic ordering in the system.
In Fig. \ref{bandas}, the horizontal dashed line at -1.5 eV and 1 eV  wrap 
the chosen energy window where the \textit{wannierization} process takes place.
Within the selected energy window, the number of desired Wannier functions is equal 
to the number of $t_{2g}$-orbitals  multiplied by the number Cr of atoms in the unit 
cell, which in our case is 12 (3 $t_{2g}$-orbitals times 4 Cr atoms per unit cell).
Also, we have to take into account that there are 14 bands lying inside the 
chosen energy window, meaning that the disentanglement procedure must be used 
before the \textit{wannierization} procedure takes place.
Nevertheless, fast and accurate \textit{wannierization}  can be done 
using as initial guess the expected $t_{2g}$-like wave functions. 
Our results fulfill the convergence criteria established in Ref. 
\onlinecite{wannier90}. The tight binding fit, if included in Fig. \ref{bandas},
would be indistinguishable from the DFT results and therefore we omit them for
the sake of clarity

\begin{figure}[t]
\centering
\includegraphics[scale=0.62]{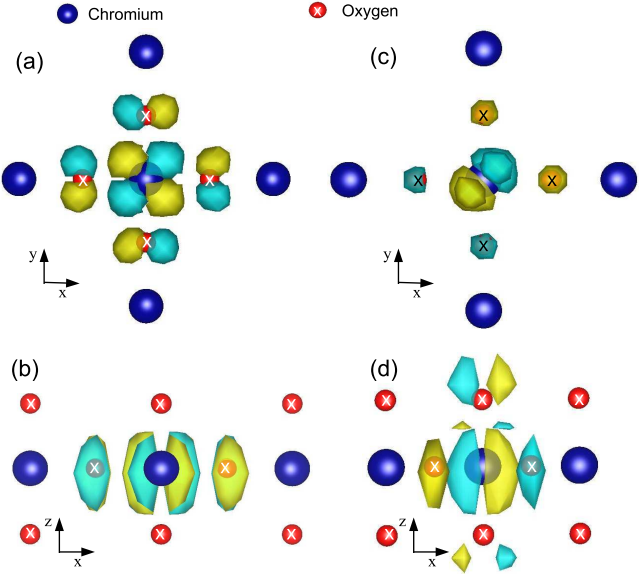}
\caption{(Color online) Isosurface plot of two maximally localized Wannier wave functions 
center at the Cr atom (blue circle). a) and b) shows two views of the 
wave function with $xy$ symmetry, showing strong hybridization with in plane 
oxygens (O$_1$). c) and d) shows two views of the  wave function
with symmetry near $xz+yz$.  Note in d) the different  
hybridization between oxygen on top of chromium (O$_2$) and the one at the 
bottom (O$_3$). 
The Sr atoms are not shown.}
\label{WF1} 
\end{figure}

The obtained MLWFs confirm the expected $t_{2g}$-type symmetry and are 
centered at Cr atoms as shown in Fig. \ref{WF1}.
Nevertheless, the isosurface plot in real space of the MLWFs reveals the strong 
hybridization between Cr and O orbitals, in agreement with the projected  
density of states shown in Fig.\ref{bandas}.
The amount of covalency that the MLWFs represents can be estimated by 
integrating the projected density of states of orbitals with different 
symmetry inside the chosen energy window where the \textit{wannierization} 
procedure has been performed.
We obtain that 78.4 \% of the states correspond to Cr $t_{2g}$ orbitals,
2.3 \% to Cr $e_{g}$ orbitals and 19.3 \% to O $p$ orbitals, where the majority 
of them (14.6 \%) are $p_x$ and $p_y$ orbitals.
In addition, note that the point group at the Cr atoms does not contain 
the reflection  through the CrO$_2$ layers (in particular the distances between 
O2 and O3 atoms and the Cr atoms, $d_2$ and $d_3$ in Fig. \ref{estructura} are 
different), 
a fact that is evident in the different content of O$_2$ and O$_3$
$p$ orbitals in the Wannier functions with approximate $xz$ and $yz$ symmetry.
The aforementioned facts allows us to conclude that the hoppings obtained from 
the Hamiltonian in the 
the base of MLWFs includes the effect of O atoms in the effective Cr-Cr 
hopping processes.

This suggests that a simplified multiband model can be proposed (as we do in the 
following Section) without the need to include crossed terms between $d$ and 
$p$ orbitals.

\section{The multiband Hubbard model}

\label{h6b}

The multiband model is constructed from effective $t_{2g}$ orbitals at each
Cr site,  the difference between on-site energies and the effective hopping between orbitals at
different sites calculated with DFT+MLWFs, and the interactions between  
electrons at the same site. The Hamiltonian is 
\begin{equation}
H_{m}=\sum_{i,\mathbf{r}}(H_{CF}^{i\mathbf{r}}+H_{I}^{i\mathbf{r}})+H_{h}\;.
\label{hm}
\end{equation}%
Here $i=1,2$ indicates the upper or lower square sublattice at positions $%
z=\pm 3.87$\AA /2, and $\mathbf{r}=a(n\mathbf{\hat{x}}+m\mathbf{\hat{y}})$,
with $a=3.796$\AA\ and  $n,m$ integers,  denotes the position of a Cr atom
within the plane. The creation operators of effective  $t_{2g}$ orbitals at
site $i,\mathbf{r}$ with spin $\sigma $ are denoted by $d_{i\mathbf{r}\alpha
\sigma }^{\dagger }$, where $\alpha =xy$, $xz$, or $yz$. 
As discussed in the previous Section, these effective orbitals contain some admixture 
with 2p O orbitals (and probably other orbitals), 
and do not have a definite symmetry under the reflection 
$z \rightarrow -z$. However, for simplicity we keep the notation 
corresponding to $t_{2g}$ orbitals in cubic symmetry ($xy$, $xz$, or $yz$).

The first term in
Eq. (\ref{hm}) corresponds to the tetragonal crystal field, which raises the
energy of an electron in the $d_{xy}$ orbital with respect to the other two.

\begin{equation}
H_{CF}^{i\mathbf{r}}=-\delta \sum_{\sigma }n_{i\mathbf{r}xy\sigma },
\label{hcf}
\end{equation}%
where $n_{i\mathbf{r}\alpha \sigma }=d_{i\mathbf{r}\alpha \sigma }^{\dagger
}d_{i\mathbf{r}\alpha \sigma }$. The second term of $H_{m}$ contains the
on-site interactions and takes the form \cite{ruo,spli,superc}

\begin{eqnarray}
H_{I}^{i\mathbf{r}} &=&U\sum_{\alpha }n_{i\mathbf{r}\alpha \uparrow }n_{i%
\mathbf{r}\alpha \downarrow }+\frac{1}{2}\sum_{\alpha \neq \beta ,\sigma
\sigma ^{\prime }}(U^{\prime }n_{i\mathbf{r}\alpha \sigma }n_{i\mathbf{r}%
\beta \sigma ^{\prime }}  \notag \\
&&+Jd_{i\mathbf{r}\alpha \sigma }^{\dagger }d_{i\mathbf{r}\beta \sigma
^{\prime }}^{\dagger }d_{i\mathbf{r}\alpha \sigma ^{\prime }}d_{i\mathbf{r}%
\beta \sigma })  \notag \\
&&+J\sum_{\alpha \neq \beta }d_{i\mathbf{r}\alpha \uparrow }^{\dagger }d_{i%
\mathbf{r}\alpha \downarrow }^{\dagger }d_{i\mathbf{r}\beta \downarrow }d_{i%
\mathbf{r}\beta \uparrow }.  \label{hi}
\end{eqnarray}%
In the following we will take $U^{\prime }=U-2J$ that corresponds to
spherical symmetry \cite{spli} (as for the free atom). 

The hopping between  $t_{2g}$ orbitals is mediated by Cr-O hopping through O
2p orbitals and the symmetry of the orbitals imposes restrictions on the
allowed processes. As a consequence, the  $xy$ orbitals cannot
hop in the $z$ direction. Similarly the $xz$ ($yz$) orbitals
cannot hop in the $y$ ($x$) direction. Then, the hopping term of the multiband
model has the form

\begin{eqnarray}
H_{h} &=&[-t_{z}\sum_{\mathbf{r}}\sum_{\alpha \neq xy}d_{1\mathbf{r}\alpha
\sigma }^{\dagger }d_{2\mathbf{r}\alpha \sigma }-t_{xy}\sum_{i,\mathbf{r}%
}d_{i\mathbf{r+a,}xy\sigma }^{\dagger }d_{i\mathbf{r,}xy\sigma }  \notag \\
&&-t_{p}\sum_{i,\mathbf{r}}(d_{i\mathbf{r\mathbf{+}a\mathbf{\hat{x}},}%
xz\sigma }^{\dagger }d_{i\mathbf{r,}xz\sigma }+d_{i\mathbf{r+}a\mathbf{\hat{y%
},}yz\sigma }^{\dagger }d_{i\mathbf{r,}yz\sigma })  \notag \\
&&+\text{H.c.}],  \label{hh}
\end{eqnarray}%
where $\mathbf{a}$ is a vector connecting two nearest Cr atoms in the $+x$
or $+y$ direction.

The crystal-field splitting  $\delta$ and the hopping parameters 
$t_{z}$, $t_{p}$ and $t_{xy}$
were determined from the MLWFs, as described in Section  \ref{hopar}.

The resulting crystal-field splitting is $\delta =0.040$ eV. This is likely
an underestimation, because orbital polarization, which is not properly
taken into account by DFT leads to larger splittings.\cite{spli} In any
case, its detailed value affects very little our results and does not change
our conclusions as long as $\delta >0$. A negative $\delta $ would lead to a
trivial pseudospin singlet configuration $d_{xz}^{1}d_{yz}^{1}$ at each site
and is inconsistent with a change of entropy near $R\ln (6)$ observed in the
magnetic transition.\cite{srcro} 

The resulting hopping parameters are $t_{z}=0.235$ eV, $t_{p}=0.214$ eV and $%
t_{xy}=0.248$ eV. They are of the same order of magnitude. The fact that $%
t_{xy}>t_{p}$ is expected in perturbation theory in the Cr-O hopping because
in the expression for  $t_{xy}$, the denominator involves the energy
necessary to take an electron from the O atom and put it in an $xy$ orbital,
and this is smaller than the corresponding denominator for $t_{p}$. Instead,
the fact that $t_{z}>t_{p}$ is rather unexpected, because the interlayer
distance is almost 2\% larger than $a$, the shortest distance between Cr
atoms in the plane. \ However, different on-site energies of the O\ orbitals
lying in between the Cr atoms and the fact that the effective orbitals are
deformed with respect to the ideal shape (as evidenced in Fig. \ref{WF1}) 
can modify the result. The deformation of the
orbitals can also explain an effective hopping between  $xz$ ($yz$) orbitals
in the $y$ ($x$) direction of magnitude 0.035 eV absent in perturbation
theory  in the Cr-O hopping because of symmetry. We neglect this
contribution.

Concerning the values of the interactions $U$ and $J$,  a fit of the lowest
atomic-energy levels along the 3d series gives $J=0.70$ eV.\cite{kroll} We
take this value as a basis for our study. Since $J$ does not involve charge
transfer, it is usually not screened in the solids in contrast to $U$. 
However, in our case since the effective $t_{2g}$ orbitals contain some
orbital admixture, $J$ can be smaller. 
A reasonable value for $U$ for early transition metals, already used to study 
the 
orbital Kondo effect in V-doped 1{\it T}-CrSe$_2$,\cite{kov} 
is $\sim 4$ eV. We shall analyze the dependence of the results with $U$.

\section{The Kugel-Khomskii Hamiltonian}

\label{hkk}

\bigskip For two electrons per site and large enough $U$ the system
described by the multiband Hamiltonian Eq. (\ref{hm}) leads to an insulating
ground state, in agreement with the experimental evidence in 
Sr$_{3}$Cr$_{2}$O$_{7}$.\cite{srcro} In this case, the hopping term $H_{h}$ can be
eliminated from $H_{m}$ by means of a canonical transformation 
(similarly to the derivation of the Heisenberg model
from the Hubbard one \cite{heis}) leading to an effective Kugel-Khomskii
Hamiltonian for the spin 1 and pseudospin 1/2 (orbital) degrees of freedom.
The eigenstates and corresponding energies of $H_{m}-H_{h}$ that we need
[see Eqs. (\ref{hm}), (\ref{hcf}) and (\ref{hi})] have been calculated in 
Ref. \onlinecite{ruo}. We restrict 
to
second order in $H_{h}$ and denote the spin at each site by $\mathbf{S}_{i%
\mathbf{r}}$ and the pseudospin by $\mathbf{T}_{i\mathbf{r}}$, with $T_{i%
\mathbf{r}}^{z}=-1/2$ (1/2) corresponding to the $d_{xy}^{1}d_{xz}^{1}$ ($%
d_{xy}^{1}d_{yz}^{1}$) configuration. The resulting effective Hamiltonian
can be written as 
\begin{equation}
H_{\text{KK}}=\sum_{\mathbf{r}}H_{\mathbf{r}}^{\perp }+\sum_{i}H_{i}^{p},
\label{hst}
\end{equation}%
where $H_{\mathbf{r}}^{\perp }$ contains the vertical interactions (in the $z
$ direction) for each two-dimensional position $\mathbf{r}$ in the $xy$
plane, and $H_{i}^{p}$ describes the interactions in the plane $i=1$, 2.
Dropping irrelevant constants, one has

\begin{eqnarray}
H_{\mathbf{r}}^{\perp } &=&\frac{I_{S}}{4}\mathbf{S}_{1\mathbf{r}}\cdot 
\mathbf{S}_{2\mathbf{r}}+I_{T}\mathbf{T}_{1\mathbf{r}}\cdot \mathbf{T}_{2%
\mathbf{r}}  \notag \\
&&+I_{ST}(\mathbf{S}_{1\mathbf{r}}\cdot \mathbf{S}_{2\mathbf{r}})(\mathbf{T}%
_{1\mathbf{r}}\cdot \mathbf{T}_{2\mathbf{r}}),  \label{hperp}
\end{eqnarray}%
where the factor 1/4 in the first term is introduced to compensate for
factors $\pm 1/4$ that come from $\mathbf{T}_{1\mathbf{r}}\cdot \mathbf{T}_{2%
\mathbf{r}}$ in classical orderings and render easier the qualitative
discussion below. The coefficients are

\begin{eqnarray}
\frac{I_{S}}{t_{z}^{2}} &=&-\frac{4}{3U_{0}}+\frac{7}{3U_{3}}+\frac{1}{U_{5}}%
,  \notag \\
\frac{I_{T}}{t_{z}^{2}} &=&\frac{8}{3U_{0}}+\frac{1}{3U_{3}}-\frac{1}{U_{5}},
\notag \\
\frac{I_{ST}}{t_{z}^{2}} &=&\frac{4}{3U_{0}}-\frac{1}{3U_{3}}+\frac{1}{U_{5}}%
,  \label{iperp}
\end{eqnarray}%
with 

\begin{eqnarray}
U_{n} &=&U_{0}+nJ,  \notag \\
U_{0} &=&U-3J.  \label{un}
\end{eqnarray}%
$U_{0}$ is the energy necessary to take a $d_{xz}$ ($d_{yz}$) electron from
the ground state of the $d_{xy}^{1}d_{xz}^{1}$ ($d_{xy}^{1}d_{yz}^{1}$)
configuration and add it to the $d_{xy}^{1}d_{yz}^{1}$ ($d_{xy}^{1}d_{xz}^{1}
$) configuration of a neighboring site to build the ground state of the $%
d_{xy}^{1}d_{xz}^{1}d_{yz}^{1}$ configuration.

Similarly for the interactions in each plane

\begin{eqnarray}
H^p_{i} &=&\sum_{\mathbf{ra}}[\frac{I_{S}^{p}}{4}\mathbf{S}_{i\mathbf{r}}\cdot 
\mathbf{S}_{i\mathbf{r}+\mathbf{a}}+I_{T}^{p}T_{i\mathbf{r}}^{z}T_{i\mathbf{r%
}+\mathbf{a}}^{z}  \notag \\
&&+I_{ST}^{p}(\mathbf{S}_{j}\cdot \mathbf{S}_{j+\mathbf{a}})T_{i\mathbf{r}%
}^{z}T_{i\mathbf{r}+\mathbf{a}}^{z}]  \notag \\
&&+I_{A}\sum_{\mathbf{r}}[-(\mathbf{S}_{i\mathbf{r}}\cdot 
\mathbf{S}_{i\mathbf{r+}a\mathbf{\hat{x}}})(T_{i\mathbf{r}}^{z}+T_{i\mathbf{r+}
a\mathbf{%
\hat{x}}}^{z})  \notag \\
&&+(\mathbf{S}_{i\mathbf{r}}\cdot \mathbf{S}_{i\mathbf{r+}a\mathbf{\hat{y}}%
})(T_{i\mathbf{r}}^{z}+T_{i\mathbf{r+}a\mathbf{\hat{y}}}^{z})],  \label{hp}
\end{eqnarray}%
where

\begin{eqnarray}
I_{S}^{p} &=&I_{xy}+t_{p}^{2}\left( -\frac{2}{3U_{0}}+\frac{7}{6U_{3}}+\frac{%
1}{2U_{5}}\right) ,  \notag \\
\frac{I_{xy}}{t_{xy}^{2}} &=&2\left( \frac{1+\delta /r}{U_{4}-r}+\frac{%
1-\delta /r}{U_{4}+r}\right) ,\text{ }r=\sqrt{\delta ^{2}+J^{2}}  \notag \\
\frac{I_{T}^{p}}{t_{p}^{2}} &=&\frac{4}{3U_{0}}+\frac{1}{6U_{3}}-\frac{1}{%
2U_{5}},  \notag \\
\frac{I_{ST}^{p}}{t_{p}^{2}} &=&\frac{2}{3U_{0}}-\frac{1}{6U_{3}}+\frac{1}{%
2U_{5}},  \notag \\
\frac{I_{A}}{t_{p}^{2}} &=&\frac{1}{2U_{3}}+\frac{1}{2U_{5}}.
\label{iplane}
\end{eqnarray}

At this point we discuss qualitatively the meaning of $H_{\text{KK}}$ and
the expected physics. We begin discussing the two-site vertical interactions
$H_{\mathbf{r}}^{\perp }$ [Eq. (\ref{hperp})]. For $J=0$, all interactions
are equal [see Eq. (\ref{iperp})]:  $I_{S}=I_{T}=I_{ST}=I=2t_{z}^{2}/U_{0}$.
This means that without the spin-pseudospin interaction $I_{ST}$ both spins
and pseudospin minimize the energy for an antiferromagmetic (AF) alignment,
but the term in $I_{ST}$ is minimized for one ferromagnetic (FM) and the
other AF alignment. As a consequence from the four classical possibilities
of orienting the spin and pseudospin FM or AF, all of them are part of the
degenerate ground state with energy -$I/2$ except the FM-FM\ one. This
result is easy to understand: the second order correction to the energy of
these states contains virtual processes in which one electron in the 
$xz$ (pseudospin $\downarrow $) or $yz$ (pseudospin $\uparrow $) orbital and spin 
$\uparrow $ or $\downarrow $ jumps to the other site and comes back. The
corresponding gain in energy is the same for any alignment of spin and
pseudospin except in the case in which the same orbital with the same spin
is occupied at both sites because of Pauli principle. If the 
$xy$ orbitals were absent, leaving spins 1/2, this picture would not be
modified by quantum fluctuations. Actually in this case the model would have
SU(4) symmetry with spin and pseudospin playing a similar role.\cite{fepc2}
In our actual case with $S=1$, the pseudospins 1/2 are more quantum than the
spins 1 and the ground state of the dimer is a pseudospin singlet and spin
triplet with energy $(I_{S}-3I_{T}-3I_{ST})/4=-5I/4$. The first excited
state is a pseudospin triplet and spin singlet with energy $%
(-2I_{S}+I_{T}-2I_{ST})/4$ $=-3I/4$. 

When $J$ (the interaction responsible for the Hund rules)  is increased, as
expected the ferromagnetic spin interactions are favored. From Eqs. (\ref%
{iperp}) it is apparent that $I_{S}$ decreases more strongly than the other
two, clearly favoring the pseudospin singlet and spin triplet. A disadvantage
of the pseudospin singlet is that it cannot take advantage of the pseudospin
interactions in the plane (except for some fluctuations).

Leaving aside for the moment the contribution $I_{xy}$ due to the hopping of
the $d_{xy}$ orbitals, the interactions in the plane $I_{S}^{p}$, $I_{T}^{p}$, 
$I_{ST}^{p}$ are exactly half of the corresponding ones in the vertical
direction if $t_{p}=t_{z}$. This factor  is due to the fact that 
for a given direction in the plane, only one of
the degenerate $xz$, $yz$ orbitals can hop. The anisotropy in direction is reflected by the term
proportional to $I_{A}$. Another consequence of the fact that  
$xz$ ($yz$) orbitals can only hop in the $x$ ($y$) direction in the plane is the
absence of pseudospin flip terms in $H_{k}$ [see Eq. (\ref{hp})].

For  $t_{xy}=J=0$ one has 
$I_{S}^{p}=I_{T}^{p}=I_{ST}^{p}=I_{A}=t_{p}^{2}/U_{0}$. From the influence of 
$J$ on the parameters (similar to the case of the vertical interaction) one
would expect AF pseudospin ordering and FM\ spin ordering favored. However
the contribution due to the hopping of the $d_{xy}$ orbitals dominates the
spin ordering. For $J=0$, $I_{xy}=4t_{p}^{2}/(U_{0}-\delta )$. The prefactor
4 with respect to the other interactions in the plane is due to a factor 2
because the $d_{xy}$ orbitals an hop in both the $x$ and $y$ directions, and
another factor 2 because the FM spin alignment cannot gain energy even for
AF pseudospin ordering. For hoppings of the same order of magnitude clearly 
$I_{xy}$ 
dominates over the other interactions and one expects AF spin
ordering within the planes. From the argument given above, one expects in
addition FM spin ordering between planes, in agreement with the spin
ordering observed by neutron scattering and calculated 
with \textit{ab initio} methods.\cite{srcro}

Concerning pseudospin ordering, from the values of $I_{T}$ and $I_{T}^{p}$ 
given above and results in the literature \cite{sand,wang}  for $S=0$
(neglecting spins) one would expect the quantum phase transition between an
AF Neel ordered phase and the phase with vertical singlet dimers to take
place for  $t_{z}^{2}/t_{p}^{2}\approx 2.522/2$ or $t_{z}/t_{p}=1.123$.
However for the actual spin $S=1$ of the system, the observed spin ordering
and the effect of the interactions between spins and pseudospins $I_{ST}$
and $I_{ST}^{p}$, favor singlet ordering between planes and weakens the
effective intraplane AF pseudospin interaction.   

Taking $J=0.7$ eV $U=4.1$ eV, which implies 
$U_0=2$ eV, leads to the values tabulated in Table \ref{tab1} for the
parameters of $H_{\text{KK}}$.

\begin{table}[h]
\caption{Parameters of $H_{\text{KK}}$ in meV for $U=4.1$ eV, $J=0.7$ eV,
and other parameters determined by the \textit{ab initio} calculations}
\label{tab1}%
\begin{ruledtabular}
\begin{tabular}{lllllll}
$I_S$ & $I_T$ & $I_{ST}$ & $I_{S}^{p}$ & $I_{T}^{p}$ & $I_{ST}^{p}$ & $I_{A}$  
\\ 
\hline
4.2 & 68.1 & 42.4 & 54.7 & 28.2 & 17.6 & 9.7 \\

\end{tabular}
\end{ruledtabular}
\end{table}

From the results (rather expected from the above discussion), it is clear that 
the dominant 
vertical interaction is $I_T$ which favors pseudospin singlets, or possibly AF 
vertical order. 
This together with the effect of $I_{ST}$, which is about ten times larger than 
$I_S$, overcomes
the weak antiferromagnetic interaction $I_S$ and FM vertical spin alignment is 
clearly favored.
Instead, in the planes the dominant interaction is $I_{S}^{p}$ which favors spin AF 
order.
In addition the interaction between spins and pseudospins $I_{ST}^{p}$ is 
smaller than
$I_{T}^{p}$ and therefore AF orbital order in the plane is also expected. 
Finally the
anisotropic interaction $I_A$ which favors FM orbital order is clearly smaller 
and has no relevant 
effect. 

It is interesting to note that for the related insulating compound BaCrO$_3$,
calculations using DFT and dynamical mean-field theory lead to AF
spin and orbital ordering in the CrO$_2$ planes.\cite{giova}

A detailed study of the competition between vertical pseudospin singlets and 
long-range 
pseudospin AF ordering is the subject of the following Section.

\section{The spin and pseudospin ordering}

\label{spso}

In this Section, we report on our study of the stability of two phases I and
II, which are the most likely according to the analysis of the previous
section and a numerical study on a small cluster.\cite{srcro} In both of
them, the spin ordering corresponds to the experimentally observed one:
antiferromagnetic in the planes and ferromagnetic between planes. In phase
I, the pseudospins form vertical dimer singlets, and in phase II they order
antiferromagnetically in both directions.

To calculate the energy and stability of phase I, we used the idea of the
bond-operator formalism,\cite{chubu,sach,gopa,matsu,bond} but in the form
of a generalized spin-wave theory,\cite{muniz} which allows us to avoid the
use of Lagrange multipliers. For phase II we use ordinary spin-wave theory.

The vertical pseudospin singlet $(|\uparrow \downarrow \rangle -|\downarrow
\uparrow \rangle )/\sqrt{2}$, where the first arrow denotes $T_{1\mathbf{r}}^{z}$,
can be represented using a boson operator $s_{\mathbf{r}}^{\dagger}$ 
as $s_{\mathbf{r}}^{\dagger}|0\rangle $, where $|0\rangle $ 
represents the boson vacuum.
The interplane term in the Hamiltonian mixes this state with
the triplet with projection 0, which can be represented as $t_{\mathbf{r}}^{\dagger }|0\rangle
=(|\uparrow \downarrow \rangle +|\downarrow \uparrow \rangle )/\sqrt{2}$,
because $T_{1\mathbf{r}}^{z}s_{\mathbf{r}}^{\dagger }|0\rangle =t_{\mathbf{r}%
}^{\dagger }|0\rangle /2$, $T_{2\mathbf{r}}^{z}s_{\mathbf{r}}^{\dagger
}|0\rangle =-t_{\mathbf{r}}^{\dagger }|0\rangle /2$, and the same
interchanging $s_{\mathbf{r}}$ and $t_{\mathbf{r}}$. Following Ref. 
\onlinecite{muniz} we assume for phase I that the number of triplet excitations 
is
small, and "condense" the singlets using 

\begin{equation}
s_{\mathbf{r}}^{\dagger }=s_{\mathbf{r}}=\sqrt{1-t_{\mathbf{r}}^{\dagger }t_{%
\mathbf{r}}}.  \label{cond}
\end{equation}%
For the spins we use the usual Holstein--Primakoff bosons proceeding in a
similar way.\cite{muniz} Performing a rotation of the spins in half of the
sites by $\pi $ around the $x$ axis to convert the AF order in the plane in
a translationally invariant FM order \cite{mart} and retaining as usual
terms up to quadratic in the bosonic operators, the Hamiltonian $H_{\text{KK}%
}$ [see Eqs. (\ref{hst}), (\ref{hperp}) and  (\ref{hp})] for phase I becomes

\begin{eqnarray}
H_{\text{KK}}(\text{I}) &\simeq &N\left[ I_{S}-3(I_{T}+I_{ST})-4I_{S}^{p}%
\right] /4  \notag \\
&&+\sum_{\mathbf{r}}(I_{T}+I_{ST})t_{\mathbf{r}}^{\dagger }t_{\mathbf{r}} 
\notag \\
&&+\sum_{\mathbf{r,a}}\left( \frac{I_{T}^{p}-I_{ST}^{p}}{2}\right) \left( t_{%
\mathbf{r}}^{\dagger }t_{\mathbf{r}+\mathbf{a}}^{\dagger }+\text{H.c.}%
\right)   \notag \\
&&+\sum_{i,\mathbf{r}}\left( \frac{-I_{S}+3I_{ST}+4I_{S}^{p}}{4}\right) b_{i%
\mathbf{r}}^{\dagger }b_{i\mathbf{r}}  \notag \\
&&+\sum_{\mathbf{r}}\left( \frac{I_{S}-3I_{ST}}{4}\right) \left( b_{1\mathbf{%
r}}^{\dagger }b_{2\mathbf{r}}+\text{H.c.}\right)   \notag \\
&&\sum_{i,\mathbf{r,a}}\frac{I_{S}^{p}}{4}\left( b_{i\mathbf{r}}^{\dagger
}b_{i\mathbf{r}+\mathbf{a}}^{\dagger }+\text{H.c.}\right) ,  \label{h1}
\end{eqnarray}%
where $N$ is the number of sites in a plane and $b_{i\mathbf{r}}^{\dagger }$
creates a spin excitation at two-dimensional position $\mathbf{r}$ of plane $%
i$.

Diagonalizing the Hamiltonian by means of a standard Bogoliubov
transformation, The ground-state energy becomes

\begin{eqnarray}
E(\text{I}) &\simeq &N\left[ I_{S}-3(I_{T}+I_{ST})-4I_{S}^{p}\right] /4 
\notag \\
&&+\sum\limits_{k,j=1}^{3}\frac{\lambda _{k}^{j}-A_{j}}{2},  \label{e1}
\end{eqnarray}%
with

\begin{eqnarray}
\lambda _{k}^{j} &=&\sqrt{A_{j}^{2}-B_{j}^{2}\left[ \cos (k_{x}a)+\cos
(k_{y}a)\right] ^{2}},  \notag \\
A_{1} &=&I_{T}+I_{ST},\text{ }B_{1}=I_{T}^{p}-I_{ST}^{p},  \notag \\
A_{2} &=&I_{S}^{p}, \notag \\
A_{3} &=& \frac{-I_{S}+3I_{ST}+2I_{S}^{p}}{2},  \notag \\
B_{2} &=&B_{3}=I_{S}^{p}/2.  \label{ab}
\end{eqnarray}%
Note that when $2B_{1}>A_{1}$, the system becomes unstable against creation
of triplet excitations of long wavelength $k_{x},k_{y}\rightarrow 0$ and Eq.
(\ref{e1}) becomes meaningless. In general if for some parameters
the assumed pseudospin or spin arrangements become unstable, the situation 
is detected in the numerical algorithm used to calculate the 
two-dimensional integral over $(k_x,k_y)$ by the non-analyticity of some 
expression for 
small $(k_x,k_y)$. In fact, as we show below, phase I becomes unstable near
the transition to phase II (as it might be expected).

For phase II with long range spin and pseudospin ordering, a similar
treatment as above using Holstein--Primakoff bosons leads to the following
energy

\begin{eqnarray}
E(\text{II}) &\simeq &N\left[
(I_{S}-I_{T}-I_{ST})/4-I_{T}^{p}-I_{S}^{p}+I_{TS}^{p}\right]   \notag \\
&&+\sum\limits_{k,j=4}^{5}\frac{\mu _{k}^{j}-A_{j}}{2}
+\sum\limits_{k,j=6}^{7}\frac{\lambda _{k}^{j}-A_{j}}{2},  \label{e2}
\end{eqnarray}

where

\begin{eqnarray}
A_{4} &=&A_{5}=\frac{I_{T}+I_{ST}}{2}+2I_{T}^{p}-2I_{TS}^{p},  \notag \\
\mu _{k}^{j} &=&\sqrt{A_{j}^{2}-C_{j}^{2}},  \notag \\
C_{4(5)} &=&(I_{T}^{p}-I_{TS}^{p})\left[ \cos (k_{x}a)+\cos (y_{x}a)\right] 
\notag \\
&&+(-)\frac{I_{T}+I_{ST}}{2},  \notag \\
A_{6(7)} &=&\frac{-I_{S}+I_{ST}}{4}+I_{S}^{p}-I_{ST}^{p}  \notag \\
&&+(-)\frac{I_{S}-I_{ST}}{4},  \notag \\
B_{6} &=&B_{7}=\frac{I_{S}^{p}-I_{TS}^{p}}{2}  \label{ab2}
\end{eqnarray}

As a test of our procedure we have compared the energy of the two phases when 
all interactions involving spin 
are zero (this is equivalent to take $S=0$) leaving only $I_T$ and $I_T^p$. We 
obtain a transition 
between the long-range ordered phase II for small $I_T$ to the phase of 
vertical dimers I for large $I_T$ at
$I_T/I_T^p=2.947$, 17\% larger than the value near 2.522 obtained by Monte 
Carlo 
calculations.\cite{sand,wang} Thus, our approach {\em underestimates} the 
stability of phase I.

\begin{figure}[ht]
\vspace{0.5cm}
\centering
\includegraphics[width=7.5cm]{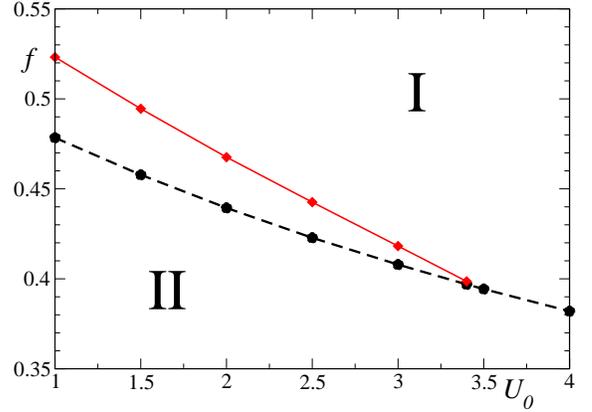}
\caption{(Color online) Factor in which the vertical hopping $t_z$ has to be 
reduced to destabilize the dimerized pseudospin
singlet phase I for $J=0.7$ eV. Full line denotes the crossing of the energies 
$E(\text{II})=E(\text{I})$ and 
dashed line is the limit of stability of phase I (see text)}
\label{f7} 
\end{figure}

For the parameters listed in Table \ref{tab1}, we obtain that the energy of the 
dimerized phase I is lower than
the long-range ordered II by 19.8 meV. These agrees with the structural 
measurements, which do not detect any distortion 
of the lattice, or displacement of the O atoms expected for long-range orbital 
ordering.
As a test of the stability of this phase, we have lowered the vertical hopping 
$t_z$ by a factor $f$ and searched for the 
value of $f$ that leads to the equality of both energies 
($E(\text{II})=E(\text{I})$) for different values of 
$U$. The results are shown by the full line in Fig. \ref{f7}. For the expected 
value of $U \sim 4.1$ eV 
($U_0=U-3J \sim 2$ eV), the resulting value of $f$ is slightly larger than the 
value of $f$ that corresponds to the 
instability of the dimerized phase against the formation of triplet excitations 
[given by $A_1=2B_1$, see the discussion after Eq. (\ref{e1})] corresponding to 
the dashed line in the figure. 
Both values of $f$ are of the order of 0.5 reflecting the fact
that the real system is far from the boundary of the phase diagram. For larger 
values of $U$, 
the dimerized phase is stabilized further. 
For $U_0 > 3.5$ eV ($U > 5.6$), the dimerized phase I becomes unstable at a 
point at which its energy 
is still lower than that of the long-range-ordered one II. This is probably a 
shortcoming of the approximations.
From the physics of the case of spin $S=0$,\cite{sand,wang} and its extension 
to two cubic sublattices,\cite{bruce}
one would expect a second-order transition between both phases and a 
coincidence of both transitions (full and dashed lines).

\begin{figure}[ht]
\vspace{0.5cm}
\centering
\includegraphics[width=7.5cm]{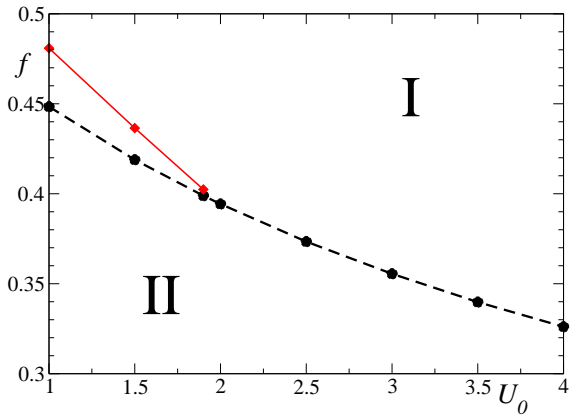}
\caption{(Color online) Same as Fig. \ref{f7} for $J=0.4$ eV.}
\label{f4} 
\end{figure}

In Fig. \ref{f4} we show how the previous results change when $J$ is reduced 
from 0.7 to 0.4 eV. We consider 
that this value is a lower bound of the interaction responsible of the Hund 
rules due to the fact that the
effective $t_{2g}$ orbitals are not pure Cr ones, but have some admixture of 
neighboring O atoms with smaller 
interactions. 
ACA In particular, from the O content estimated as described at the end of Section III,
and the fact that O sites with two $2p$ holes are very rare, 
we estimate $J=0.57$ eV.

As one can see, the changes with respect to the previous figure 
are minor. We conclude that
for the calculated values of the hopping terms obtained as described in Section \ref{hopar}, 
and reasonable 
values of the interactions, the dimerized phase I is the stable one. 

\section{Summary and discussion}
\label{sum}

Using maximally localized Wannier functions in bilayer Sr$_3$Cr$_2$O$_7$, we 
have derived a six-band Hubbard model 
for effective Cr $t_{2g}$ orbitals (three per Cr site) which contain some 
admixture with neighboring O atoms, estimated in nearly 19 \%, 
as discussed in Section III.

Using the resulting hopping and on-site energy parameters, and reasonable values 
of the 
Coulomb interaction $U$ and interaction responsible for the Hund rules $J$, we 
have derived an effective 
Kugel-Khomskii Hamiltonian $H_{\text{KK}}$ using a procedure equivalent to 
degenerate second-order 
perturbation theory in the hopping terms. 

A similar $H_{\text{KK}}$ has been derived in Ref. \onlinecite{srcro} using 
forth-order perturbation theory 
in the Cr-O hopping, but the parameters were not determined and it was not 
clear where the system lies in 
the phase diagram, although the absence of observable distortions is consistent 
with same phase that we 
obtain here.

Our analysis of $H_{\text{KK}}$ and calculations based on bond-order operators 
and spin waves,
show that the ground state of the system has long-range spin order, with 
antiferromagmetic order 
in the layers and ferromagnetic order between layers, in agreement with 
experiment and 
\textit{ab initio} calculations.\cite{srcro} Instead the orbital degrees of 
freedom 
form singlet dimers perpendicular to the planes. This rather exotic arrangement 
of the orbital degrees of freedom is rare. For interplane and intraplane 
interactions of the same order 
of magnitude, one expects long-range antiferromagmetic ordering of the 
pseudospin (orbital) degrees of 
freedom. 
Although the methods used to solve $H_{\text{KK}}$ are semiquantitative, with 
errors of the order of 17\% 
for a known case, the obtained ground state is rather far from the phase 
boundary to the phase of long-range 
antiferromagmetic pseudospin ordering.

The reason for the stability of the dimerized phase is twofold. On one-hand, 
the intraplane pseudospin interactions are smaller 
due to restrictions of the $xz$ and $yz$ orbitals to hop in 
certain directions of the plane.
On the other hand, for the spin ordering observed, 
the interactions between spins and pseudospins strengthen 
(weaken) the antiferromagmetic
pseudospin correlations normal to (in the) layers. This aspect has some 
similarities to the physics of some V oxides, for which calculations suggest that 
ferromagnetic couplings are particularly strong due to singlet orbital
fluctuations.\cite{van}

\section*{Acknowledgments}

A. A. A. (C. H.) thanks B. Normand and C. Batista (Rub\'en Weht) for helpful discussions.
A. A. A. (C. H.) is sponsored by PIP 112-201501-00506 of CONICET and PICT 2013-1045
of the ANPCyT (PICT 2014-1555 of the ANPCyT).


\begin{thebibliography}{99}
\bibitem{note} It is known that orbital-related
Coulomb interactions are underestimated in the spin LDA,\cite{esch} and
empirical orbital polarization corrections \cite{erik} are frequently used
to cure this problem.
\end{thebibliography}
\end{document}